\documentclass[useAMS,usenatbib]{mn2e}

\DeclareSymbolFont{cmletters}{OML}{cmm}{m}{it}
\DeclareMathSymbol{v}{\mathalpha}{cmletters}{"76}

\usepackage{amsmath}
\usepackage{amssymb}
\usepackage{epsfig}
\usepackage{graphicx}
\usepackage{ifthen}
\usepackage{latexsym}
\usepackage{rotating}
\usepackage{subfigure}
\usepackage{times,epsf}
\usepackage{txfonts}
\usepackage{varioref}
\usepackage{verbatim}
\usepackage{array}
\usepackage{color}
\usepackage[dvipsnames]{xcolor}
\usepackage[T1]{fontenc}

\newcolumntype{M}{>{$\vcenter\bgroup\hbox\bgroup}c<{\egroup\egroup$}}

\newcommand{\be}{\begin{equation}}
\newcommand{\ee}{\end{equation}}
\newcommand{\bea}{\begin{eqnarray}}
\newcommand{\eea}{\end{eqnarray}}

\newcommand{\sgra}{Sgr~A$^*$ }

\newcommand\apj{Astrophysical Journal}
\newcommand\apjl{Astrophysical Journal Letters}

\newcommand\aap{Astronomy \& Astrophysics}

\newcommand\nat{Nature}

\newcommand\mnras{Monthly Notices of the Royal Astronomical Society}

\newcommand\pasj{Publications of the Astronomical Society of Japan}

\title[G2 cloud bow shock]{
Location of the bow shock ahead of cloud G2 at the Galactic Center}
\author[A. S\k{a}dowski, R. Narayan, L. Sironi, F. {\"O}zel]
     {\parbox{\textwidth}{Aleksander S\k{a}dowski$^1$,
 Ramesh Narayan$^{1}$,  Lorenzo
         Sironi$^{1,2}$ and Feryal
         {\"O}zel$^{1,3}$\thanks{E-mail:
           asadowski@cfa.harvard.edu (AS); rnarayan@cfa.harvard.edu (RN); lsironi@cfa.harvard.edu
           (LS); 
fozel@email.arizona.edu (FO); }}\vspace{0.4cm}\\
        $^1$ Harvard-Smithsonian Center for Astrophysics, 60 Garden
        St., Cambridge, MA 02138, USA\\
 $^2$ NASA Einstein Postdoctoral Fellow\\
 $^3$ University of Arizona, 933 N. Cherry Avenue, Tucson, AZ 85721,
 USA}

\begin{document}

\maketitle

\label{firstpage}

\begin{abstract}
We perform detailed magnetohydrodynamic simulations of 
the gas cloud G2 interacting with the accretion flow around the Galactic
Center black hole Sgr~A$^*$. We take as our initial conditions a
steady-state, converged solution of the accretion flow obtained earlier
using the general-relativistic magnetohydrodynamic code HARM. Using
the observed parameters for the cloud's orbit, we compute the
interaction of the cloud with the ambient gas and identify the shock
structure that forms ahead of the cloud. We show that for many
configurations, the cloud front crosses orbit pericenter
7 to 9 months earlier than the center-of-mass. 
\end{abstract}


\section{Introduction}
\label{s.introduction}

\cite{gillessen+12a,gillessen+12b} discovered an object, likely a cloud of molecular gas,
called G2 that is approaching the Galactic Center on a highly
eccentric orbit. The center-of-mass of the cloud is expected to reach
pericenter, located at radius $R\approx3.0\times 10^{15}\,{\rm cm}$
from the supermassive black hole Sagittarius A$^*$ (Sgr A$^*$), 
at $t_0=2013.69$ (September 2013)\footnote{Recently, \cite{keckg2} obtained another
    orbital solution, derived from Br-gamma line astrometry and radial
    velocity measurements, 
resulting in a later
    time of closest approach ($t_0=2014.21$, March 2014), slightly closer periastron
    ($3800$ vs $4400 R_{\rm G}$), and a longer orbital period ($276$ vs
    $198$ yrs). In this work we use the orbital parameters derived by
\cite{gillessen+12b}. Because of the similarity of the orbits the
shock parameters and its location with respect to the center-of-mass are
expected to be similar.}. During this encounter, the cloud will move supersonically
through the hot accretion flow around Sgr A$^*$
\citep{sadowski+G21} and a bow shock will form ahead of it
\citep{narayan+12a}. The shock provides suitable conditions 
for particle acceleration and is likely to accelerate hot ambient
electrons to relativistic energies.

\cite{narayan+12a} and \cite{sadowski+G21} estimated the efficiency of
this acceleration and calculated the expected radio synchrotron
emission.  Depending on model parameters, they predicted radio fluxes
$\sim1-20$\,Jy at $\nu = 1.4\,\rm GHz$.  In addition, using simplified
assumptions, \cite{sadowski+G21} estimated that the bow shock will
cross pericenter roughly 5 months before the center-of-mass does,
and that the peak radio flux will be reached
roughly a month later. However, their estimate did not take into
account the structure of the cloud or the details of the dynamical
interactions between the cloud and the ambient gas, both of which have
the potential to change the shock properties and the expected radio
signals.

In this paper we obtain a more reliable estimate of the shock location
and the interaction epoch by carrying out general relativistic
magnetohydrodynamic (GRMHD) simulations of G2 moving through the
turbulent hot accretion flow around Sgr A$^*$. The simulations are
performed with the GRMHD code HARM \citep{gammieetal03} using a
realistic model of the accretion flow \citep{narayan+12b}.The 
cloud is scaled down to fit within the converged region of the disk
simulation.
Although
the spatial resolution of the simulations is limited, we are able to
determine with reasonable confidence that the bow shock is formed
ahead of the cloud, and that it crosses
pericenter 7 to 9 months earlier than $t_0$. The accompanying
radio emission is thus expected to reach peak values in Spring 
or late Summer 2013 for the \cite{gillessen+12b} and \cite{keckg2}
orbital models, respectively.

Our simulations are the first that include magnetic fields and
  propagate the cloud through a turbulent accretion flow in three
  dimensions (3D), thus probing phenomena that were not explored in
  previous studies. \cite{burkert+12} and \cite{shartmann+12}
  performed two-dimensional fixed-mesh simulations of the cloud
  approaching \sgra and penetrating an isotropic, smooth, non-magnetized
  and artificially stabilized atmosphere. Similar assumptions about
  the background gas were adopted by \cite{anninos+12} who performed
  3D adaptive mesh simulation, which are
  best suited for resolving instabilities at the cloud surface
  (however, they still neglected the role of magnetic fields). Finally,
  \cite{saitoh+12} used a Smoothed Particle Hydrodynamics (SPH) code
  to simulate the evolution of the cloud assuming the same
  model for the ambient gas as the other authors, but allowing the gas
  to cool through optically-thin radiation.

The structure of the paper is as follows. In Section~\ref{s.model} we
discuss the initial conditions for our simulations, in
Section~\ref{s.results} we describe the simulation results, and in
Section~\ref{s.discussion} we discuss uncertainties and the optimal
strategy for radio observations of the G2 impact. 

\section{Model}
\label{s.model}

\subsection{Disk model}

The radiatively inefficient accretion flow around \sgra has been
studied extensively through observations and theoretical
investigations. The best available constraints come at large radii
from the measurement of the gas density at the Bondi radius, which
yields $n_{\rm e}(R_{\rm B})\approx130\, \rm cm^{-3}$ (Baganoff et
al. 2003), and in the innermost regions, where the mass accretion rate
onto the black hole has been estimated to be $\dot M_{\rm BH}\approx
10^{-9}\pm 10^{-8}\, \rm M_\odot\,{\rm yr^{-1}}$ \citep[see
][]{yuan+03,marrone07,moscibrodzka+09,dexter+10,roman+12,dibi+12}. Models of the gas density and
the electron temperature at these small distances can simultaneously
account for the long-wavelength spectrum of the accretion flow, the
polarization measurements, and the size of the emitting region at
1.3~mm (see Broderick \& Loeb 2005). Less is known directly about the
properties of the flow at intermediate radii, where the pericenter of
G2's orbit lies. Psaltis (2012) carried out a comparison of the gas
properties obtained at intermediate radii by extrapolating the flow
properties inwards from the Bondi radius using the model of Quataert
(2004) and outwards from the model of Broderick \& Loeb (2005) and
found reasonable agreement between these solutions between few $\times
10^2$ to $\sim 10^5 R_{\rm G}$ where $R_{\rm G}=GM/c^2$ and $M$ is the mass of the black hole.

\begin{figure}
  \centering
\begin{tabular}{MM}
\begin{sideways}$y\,(\tilde R_{\rm G})$\end{sideways}&\hspace{-.46cm}
\includegraphics[width=.95\columnwidth]{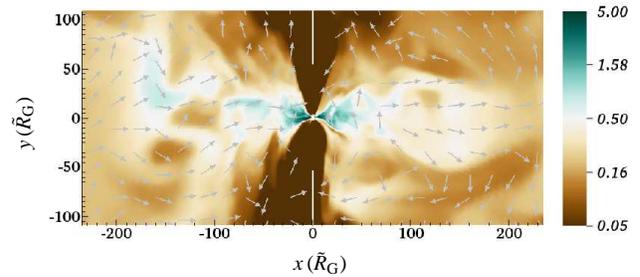}\vspace{-.1cm}\\
&\hspace{-.8cm}$x\,(\tilde R_{\rm G})$\\
\end{tabular}
\caption{Snapshot of density (arbitrary units) in the $r\theta$ plane for
the background model (evolved without the cloud) of the accretion disk 
taken at the same time as in Fig.~\ref{f.panels} (which shows the results
of cloud and disk interaction).}
\label{f.originalslice}
\end{figure}

Numerical simulations similarly rely on the measurements at small and
large radii as anchor points to calculate the properties of the gas
throughout the flow. In this paper, we make use of the radiatively
inefficient accretion simulation around a non-spinning black hole
described in \cite{narayan+12b}, which corresponds to the so-called
Magnetically Arrested Disk (MAD, Narayan et al. 2003) regime. The
  accretion disk was solved on a fixed, horizon-penetrating grid with
  264 cells logarithmicaly spaced in radius, 126 cells roughly uniform
  in $\theta$ and 60 cells uniform in
  $\phi$. It was initialized as an equilbrium torus threaded by seed,
  single-loop magnetic field \citep{penna+torus}. Magneto-Rotational
  Instability (MRI) develops, the gas accretes and it brings significant
  magnetic flux onto the BH. We adopted an ideal equation of state
  with adiabatic index $\gamma=5/3$. Fig.~\ref{f.originalslice}
shows a typical snapshot of the gas density in the $r\theta$ plane.

Because of the very long duration of this simulation, $t = 2.2\times
10^5 R_{\rm G}/c$, the simulated accretion flow is in steady state out
to a relatively large radius $\sim200R_{\rm G}$ at the equatorial
plane, and to much larger radii at higher latitudes. We normalize the
gas density in the simulation such that it is consistent with the
measured density at the Bondi radius and, at the same time quantitatively
agrees with the expected accretion rate at the BH horizon.
Shcherbakov (2013, private communication) have shown that the simulated
spectrum based on this model agrees with observed spectrum of Sgr~A$^*$.
Fitting formulae for the mean properties of the accretion flow
are given in \cite{sadowski+G21}.
The mass of the disk gas inside the cloud radius of pericenter is 
roughly $0.1M_{\rm Earth}$ and is much smaller than the mass of the
cloud $M_{\rm cl}\approx 3M_{\rm Earth}$ \citep{gillessen+12a}.

\subsection{Cloud model}

To set up the configuration of the cloud in phase space, which
we use as initial conditions for our cloud-disk interaction
simulations, we follow the approach described in \cite{gillessen+12a}.
We start with a spherical cloud of test particles at an earlier epoch
corresponding to year $2000.0$. We adjust the location and the
velocity of the center-of-mass of this initial cloud so that they are
consistent with the orbit described in \cite{gillessen+12b}. We also
assign a spherically symmetric density profile and a distribution of 
random velocities to the cloud (described in detail in the next
paragraph) and
adjust these so as to satisfy the distributions of positions and
velocities observed in year 2012 \citep{gillessen+12b}. This procedure
does not yield a unique model for the initial cloud. Therefore, in
addition to the model described in \cite{gillessen+12a}, which we
refer to as model N1, we consider a number of other models which are
all consistent with the data (see Table~\ref{t.models}).  Our aim is
to explore a range of models in order to assess the effect of the initial
cloud configuration on the location and properties of the bow shock.

In the alternative models we consider, we model the density profile
and the velocity distribution using two parametrizations for the
former and three for the latter.  For the initial density profile
(year 2000), we consider a \textit{Gaussian} distribution, with two values 
of the full width at half maximum (FWHM), and a constant density cloud (called \textit{flat}), with
three choices of the outer radius $R_{\rm out}$ (Table~\ref{t.models}). For the
\textit{Gaussian} density model, we truncate the cloud at a radius equal to
$2\sigma$, where $\sigma$ is for the standard deviation. For the velocity distribution, we take: 1. a Gaussian
distribution with FWHM $= 120~{\rm km\,s^{-1}}$ as in
\citealt{gillessen+12a} (called \textit{normal}), 2. a distribution 
where the magnitude of the additional velocity component is
proportional to the distance from the center-of-mass, with a velocity
of $120~{\rm km\,s^{-1}}$ at FWHM or $R_{\rm out}$ for \textit{Gaussian} and \textit{flat} 
models, respectively (called
\textit{scaled}), and 3. a model with no velocity dispersion, where
all the particles in the cloud move with the same velocity as the
center-of-mass (called \textit{no}). We give in Table~\ref{t.models}
the details corresponding to all the models we have considered. Each
of these models, when propagated forward from year 2000 to year 2012,
gives a cloud of particles with positions and velocities that are
consistent with observations.

\subsection{Initial MHD setup}
\label{s.mhdsetup}

The GRMHD simulation that we use to model the background accretion
flow \citep{narayan+12b} has achieved steady state out to a radius
$R\sim200R_{\rm G}$ at the equatorial plane, whereas the cloud orbit
has a pericentric radius of $4400 R_{\rm G}$.  To evolve the
cloud-disk system numerically we need to either extrapolate the disk
model to several thousand $R_{\rm G}$ or scale down the cloud and its orbit
to $\sim100R_{\rm G}$.  We choose to do the latter since it allows us to
retain the full 3D turbulent structure of the simulated disk.

We scale all distances down by a factor $X=40$, which brings the
pericenter of the orbit down to $R=110R_{\rm G}$. This choice,
although ``ad hoc'', ensures
that, on the one hand, the cloud fits within the converged region of
the simulation and, on the other hand, it does not go too close to the
black hole, where the disk model is no longer self-similar
\citep[see][]{sadowski+G21}. 
We scale the cloud size, orbit semi-major axis,
and other relevant distances by a factor of $X^{-1}$, the orbital
velocity by $X^{1/2}$, and the time relative to pericenter by
$X^{-3/2}$. The gas density in the simulation varies with radius as $R^{-1}$
\citep{sadowski+G21}. 
Therefore, we scale the cloud density also by $X^{-1}$.

The strength of the expected bow shock depends on its Mach number ---
  the ratio of the relative cloud-disk velocity to the disk sound
  speed.  The relative velocity increases proportional to the orbital velocities,
  i.e., by $X^{1/2}$.  The temperature in the numerical model follows
  $X^{1}$ and therefore the speed of sound scales as $X^{1/2}$. As a
  result, the shock strength is independent of the adopted value of
  the scaling factor $X$.

We report all simulation results in length units of $\tilde R_{\rm
G}=X GM_{Sgr A*}/c^2$, though when reporting times, e.g., in the
Tables, we convert back to the parameters of the original G2 cloud.

\begin{figure}
  \centering
\begin{tabular}{MM}
\begin{sideways}$y\,(\tilde R_{\rm G})$\end{sideways}&\hspace{-.46cm}
\includegraphics[width=.95\columnwidth]{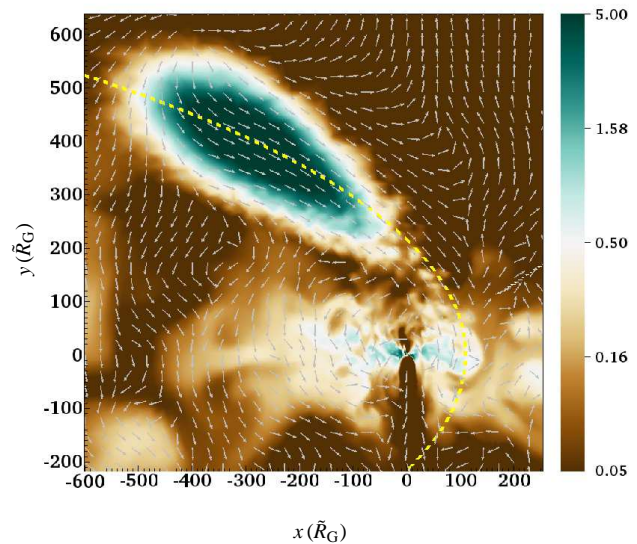}\vspace{-.1cm}\\
&\hspace{-.8cm}$x\,(\tilde R_{\rm G})$\\
\end{tabular}
\caption{Snapshot of the gas density (arbitrary units) in the $r\theta$ plane for
the initial state of the N0 model corresponding to $t=2012.5$. The dashed yellow
line shows the orbit of the cloud center-of-mass.}
\label{f.initialstate}
\end{figure}

As the initial conditions for the disk, we choose the final
configuration at time $t=2.2\times10^5R_{\rm G}/c$ of the ADAF/MAD run
described in \cite{narayan+12b}. Then, for each of the cloud models
shown in Table~\ref{t.models}, we evolve the test particles forward
along geodesics up to time $t_{\rm mhd}=2012.5$ (except for one model, N0F, where we
choose time $t_{\rm mhd}=2012.0$; see below).  We transfer the mass and momentum
corresponding to each test particle to the corresponding grid cell in
the simulation, thereby adding the cloud density and momentum to the
pre-existing accretion gas in that cell.  We assume that the cloud gas is
cold (this is certainly true relative to the $>10^8$\,K ambient gas in
the accretion disk) and unmagnetized. Thus, the pressure and magnetic
field retain the same values as in the original GRMHD simulation.
Starting from this initial state (Fig.~\ref{f.initialstate}), we run the simulation forward in
time and study the motion of the cloud as well as its interactions
with the medium.

We ran additional models to test the effects of the initial setup
on our results. We ran model N0F with a start time of 2012.0 and
confirmed that selecting a start time of 2012.5 vs. this earlier time
does not modify the results.  In order to assess the impact of the
ambient magnetic field on the cloud motion, we also ran one model
(N0B) which is identical to the fiducial model N0 except for a
magnetic field strength that we set equal to zero. We will discuss the
results of this latter run in the later sections.

\subsection{Orbit orientation}

The GRMHD simulation of the background accretion disk covers the whole
domain of $(r,\theta,\phi)$ and has sufficient resolution ($264$ grid
cells in $r$, $126$ in $\theta$, $60$ in $\phi$) to capture the
physics of the dominant MRI modes.  The cloud, however, has a
relatively small extent compared to the full domain. Therefore, the
simulations we describe here have only modest resolution on the scale
of the cloud; in particular, our resolution is coarser than other
simulations of G2 described in the literature
\citep{anninos+12,shartmann+12,saitoh+12}.

The primary effect of the numerical diffusion associated with the
coarse resolution in the simulations is under resolving
Kelvin-Helmholtz instabilities at the cloud surface. On the other
hand, the presence of magnetic fields in the external medium is
expected to dampen the Kelvin-Helmholtz instability (Markevitch \&
Vikhlinin 2007). Therefore, the potential effects of this instability
will be less significant in the present problem and in a realistic disk.

We choose an orbit orientation that allows us to utilize the highest
possible resolution to resolve the shocks that develop in the
cloud-disk interaction.  We assume that (i) the orbital plane of the
cloud is highly inclined (inclination angle $i=60$\,deg) with respect
to the equatorial plane of the accretion flow, (ii) pericenter is in
the equatorial plane (argument of periapsis $\omega=0$, corresponding
to a single disk crossing, see \citealt{sadowski+G21}), and (iii) the
cloud and the disk gas counter-rotate with respect to each other.
This ensures that the shock normal at pericenter is nearly along the
$\theta$ direction, where the simulations have the highest resolution.
As we show in the next section, the setup we use is sufficient to
reasonably resolve the bow-shock region. For comparison, we also
consider models with a co-rotating orbit (model N0C), two
low-inclination models (N0LN and N0LC) and a model with $\omega=\pi/2$
(double disk crossing, N0D).  Parameters of all the models are listed
in Table~\ref{t.models}.

\begin{table*}
\caption{Cloud models}
\label{t.models}
\centering\begin{tabular}{@{}cccccccccc}
\hline
\hline
Model& Density & Size & Velocity & $t_{\rm MHD}$ & Inclination &
Argument of & Rotation & Magnetic & Relative time of\\
name& profile &  & dispersion &                 & $i$ & periapsis $\omega$ & & field & shock pericenter\\
\hline
N0& Gaussian & FWHM=$3\times 10^{15} \rm cm$& scaled & $2012.5$ & $\pi/3$ & $0$ & counter-&  on & $-0.63$  \\ 
N1& Gaussian & FWHM=$3\times 10^{15} \rm cm$& normal & $2012.5$ & $\pi/3$ & $0$ & counter- &  on &  $-0.68$  \\ 
N2& Gaussian & FWHM=$4.5\times 10^{15} \rm cm$& no & $2012.5$ & $\pi/3$ & $0$ & counter-&  on & $-0.82$  \\ 
F0& Flat & $R_{\rm out}=2.5\times 10^{15} \rm cm$& no & $2012.5$ & $\pi/3$ & $0$ & counter-&  on & $-0.63$  \\ 
F1& Flat & $R_{\rm out}=1.5\times 10^{15} \rm cm$& scaled & $2012.5$ & $\pi/3$ & $0$  & counter-&  on & $-0.65$  \\ 
F2& Flat & $R_{\rm out}=3.5\times 10^{15} \rm cm$& no & $2012.5$ & $\pi/3$ & $0$ & counter-&  on &  $-0.92$ \\ 
N0F& Gaussian & FWHM=$3\times 10^{15} \rm cm$& scaled & $2012.0$ & $\pi/3$ & $0$ & counter-&  on &$-0.60$   \\ 
N0C& Gaussian & FWHM=$3\times 10^{15} \rm cm$& scaled & $2012.5$ & $-\pi/3$ & $0$ & co-&  on & $-0.66$  \\ 
N0LN& Gaussian & FWHM=$3\times 10^{15} \rm cm$& scaled & $2012.5$ & $\pi/6$ & $0$ & counter-&  on & $-0.54$  \\ 
N0LC& Gaussian & FWHM=$3\times 10^{15} \rm cm$& scaled & $2012.5$ & $-\pi/6$ & $0$ & co-&  on &  $-0.61$ \\ 
N0D& Gaussian & FWHM=$3\times 10^{15} \rm cm$& scaled & $2012.0$ & $\pi/3$ & $\pi/2$ & counter-&  on & $-0.54$  \\ 
N0B& Gaussian & FWHM=$3\times 10^{15} \rm cm$& scaled & $2012.5$ & $\pi/3$ & $0$ & counter-& off & $-0.70$  \\ 
& & & & & & & -rotating \\
\hline
\hline
\end{tabular}\vspace{.15cm}
\\$t_{\rm MHD}$ is the epoch when the corresponding MHD simulation
starts. The epoch of shock pericenter is given with respect to the
center-of-mass pericenter crossing. 
\end{table*}

\section{Results}
\label{s.results}

\subsection{Physical quantities near the cloud front}

\begin{figure*}
  \centering
\begin{tabular}{MMMM}
\begin{sideways}$y\,(\tilde R_{\rm G})$\end{sideways}&\hspace{-.4cm}
\includegraphics[width=.33\textwidth]{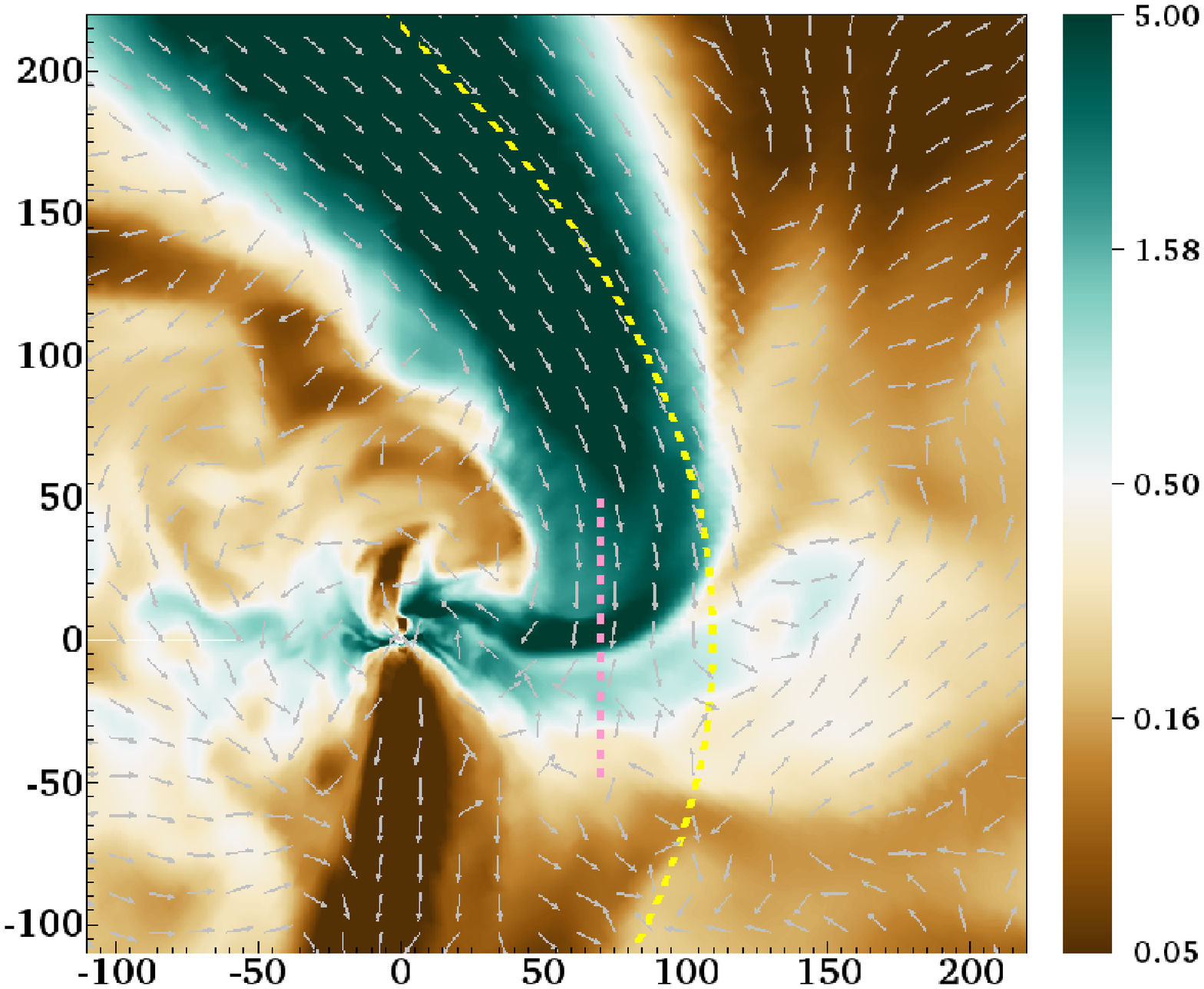}&\hspace{-.4cm}
\includegraphics[width=.33\textwidth]{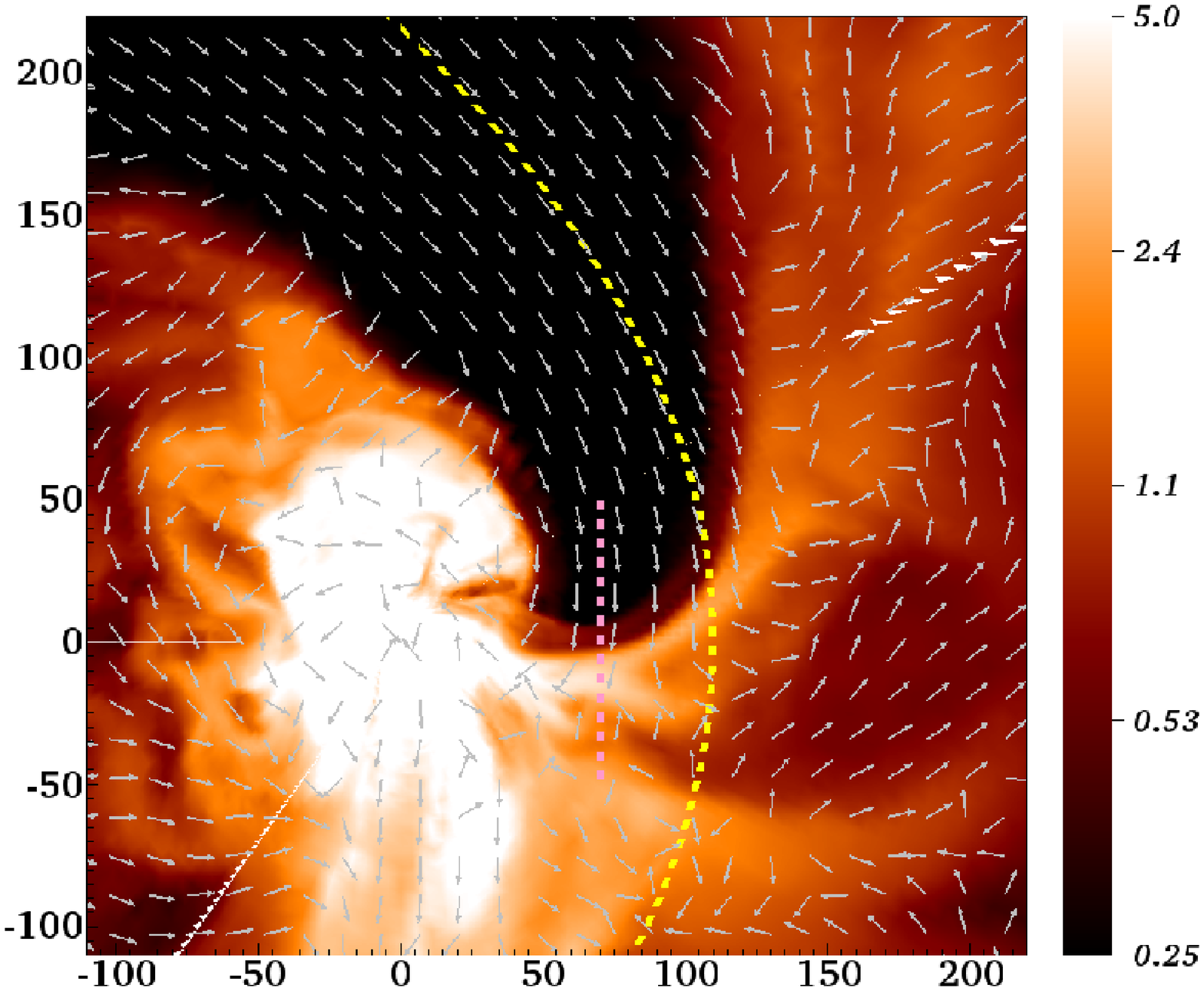}&\hspace{-.4cm}
\includegraphics[width=.33\textwidth]{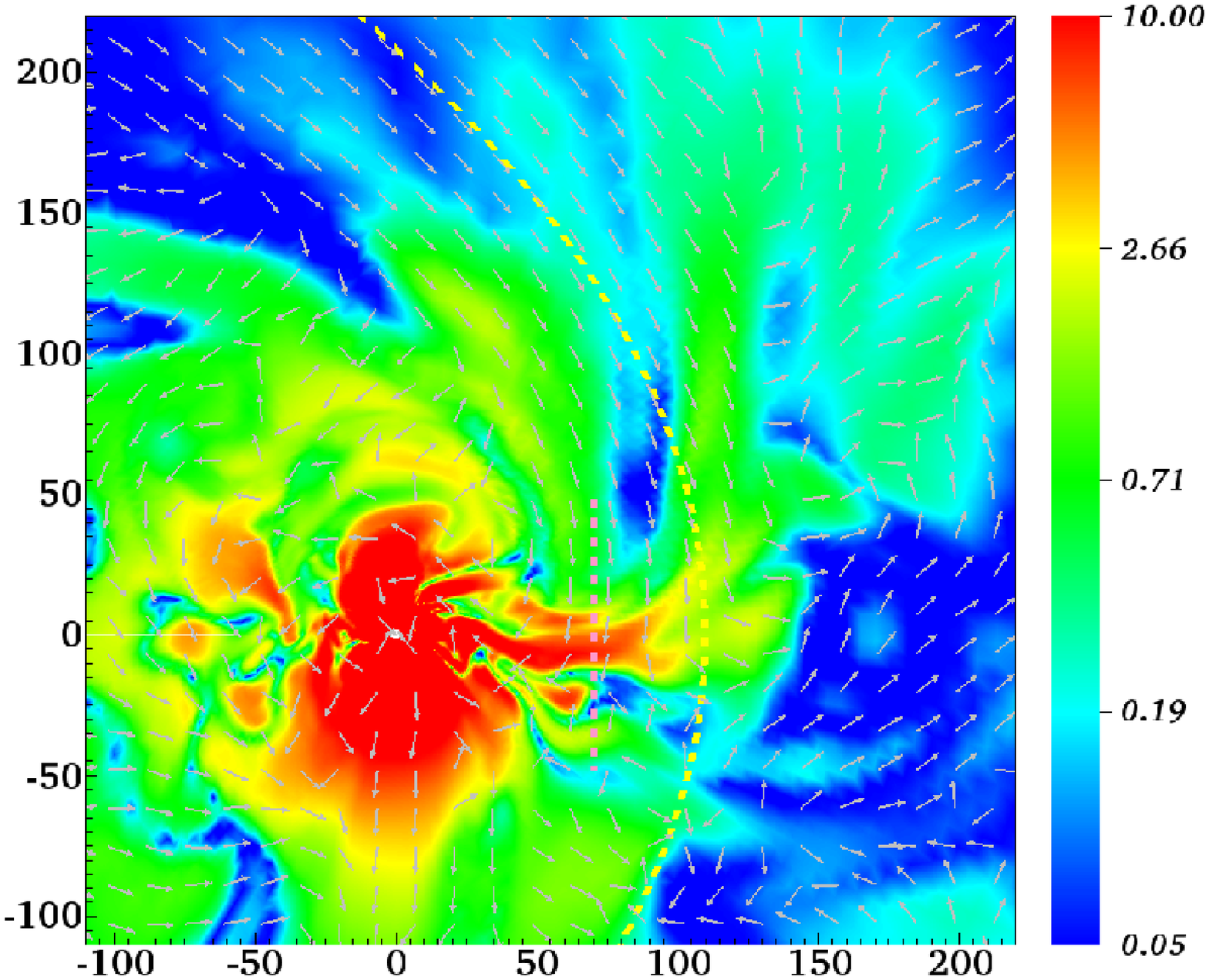}\\
\vspace{-.25cm}
\begin{sideways}$y\,(\tilde R_{\rm G})$\end{sideways}&\hspace{-.4cm}
\includegraphics[width=.33\textwidth]{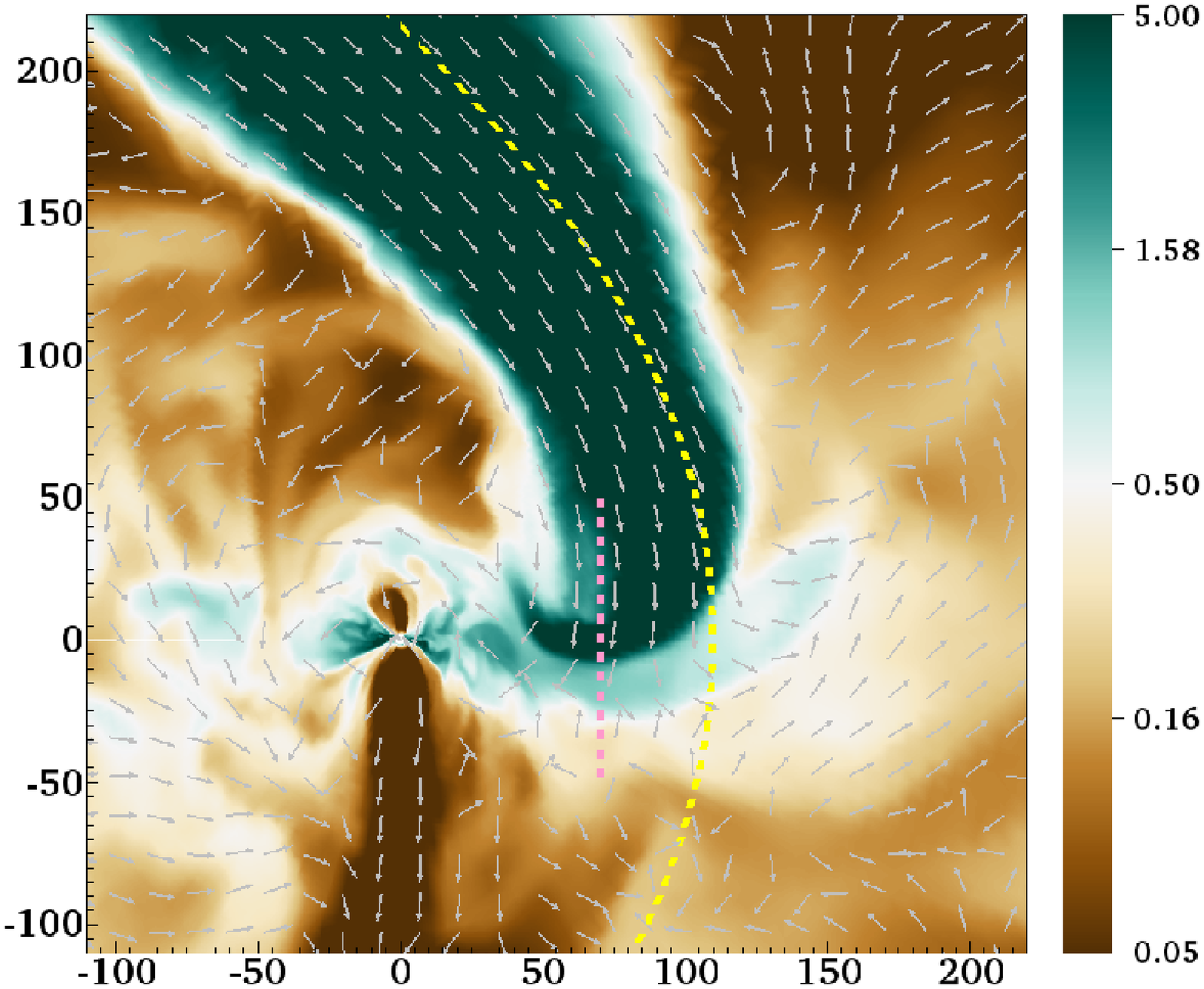}&\hspace{-.4cm}
\includegraphics[width=.33\textwidth]{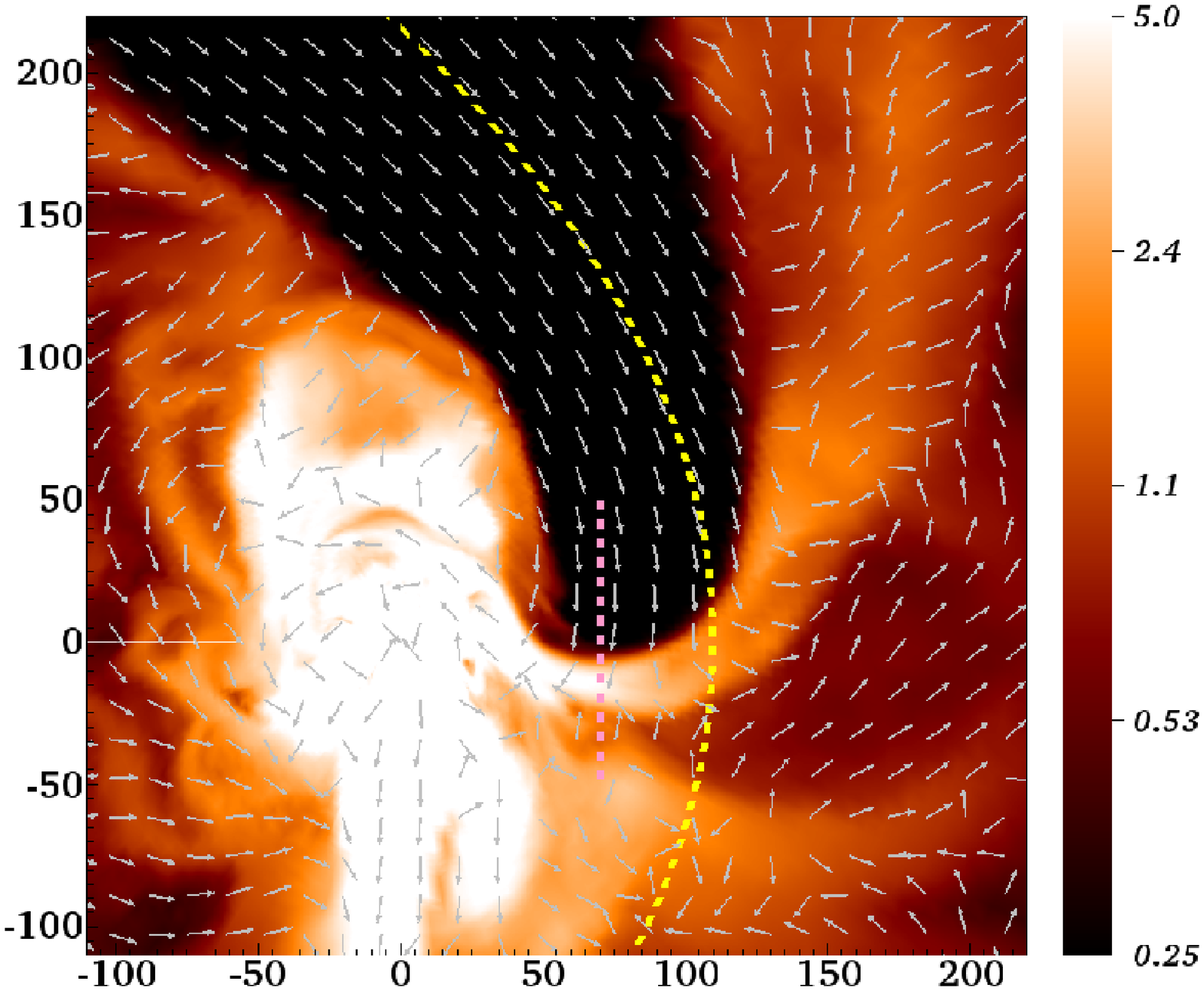}&\hspace{-.4cm}
\includegraphics[width=.33\textwidth]{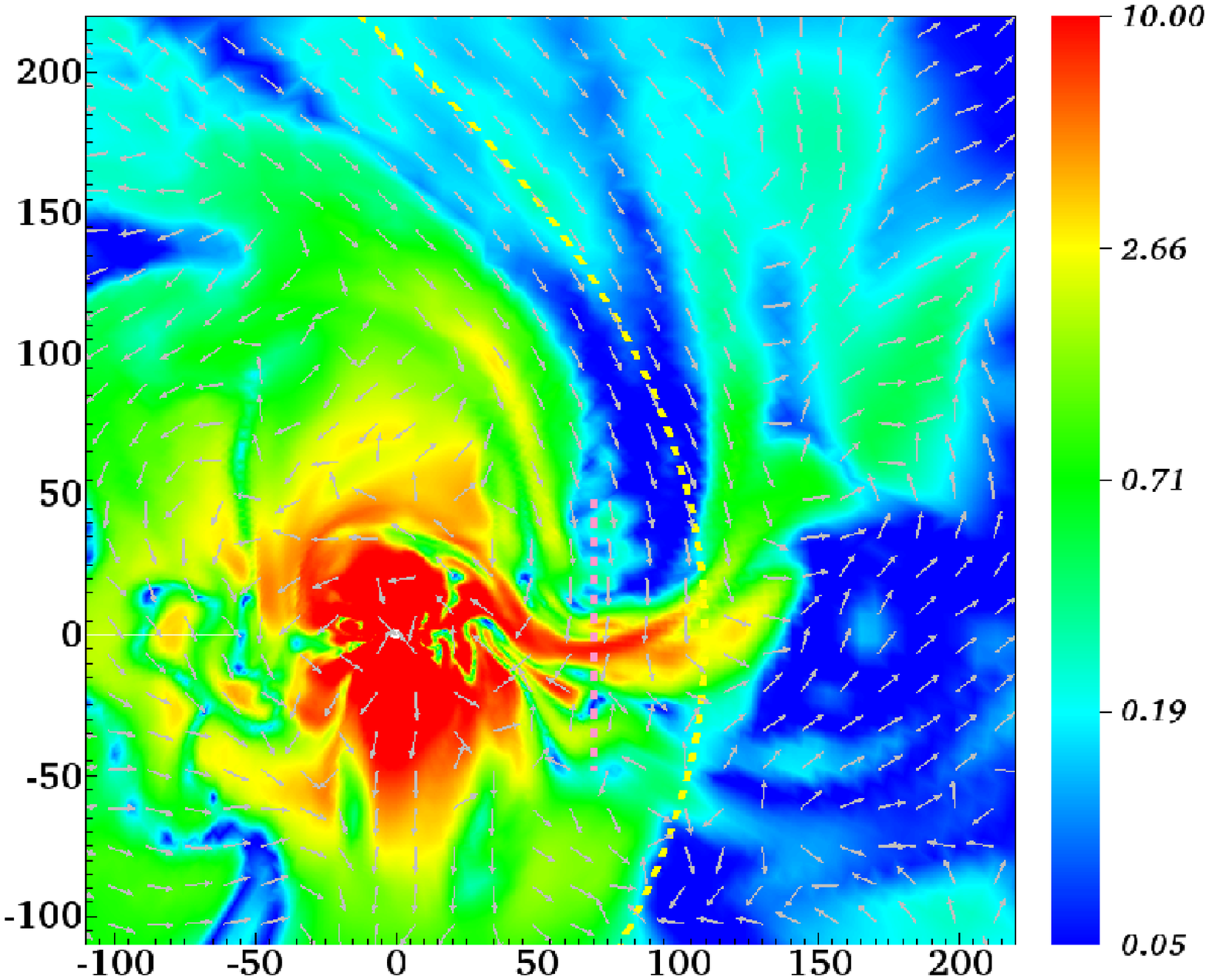}\\
&$x\,(\tilde R_{\rm G})$&$x\,(\tilde R_{\rm G})$&$x\,(\tilde R_{\rm G})$\\
\end{tabular}
\caption{Profiles of density (left), temperature (middle) and
magnetic pressure (right panel) on the cloud orbital plane at
$t=t_0-0.63\rm yr$ for model N0 (top) and F0 (bottom). Vectors show the
velocity field. The yellows dashed line is the cloud center-of-mass
orbit while the vertical lines show the cut used to extract profiles
across the shock shown in Fig.~\ref{f.lines}, \ref{f.cl2time}, and
\ref{f.cl2flattime}.}
\label{f.panels}
\end{figure*}

\begin{figure}
  \centering
\hspace{-.05cm}\includegraphics[angle=270,width=1.\columnwidth]{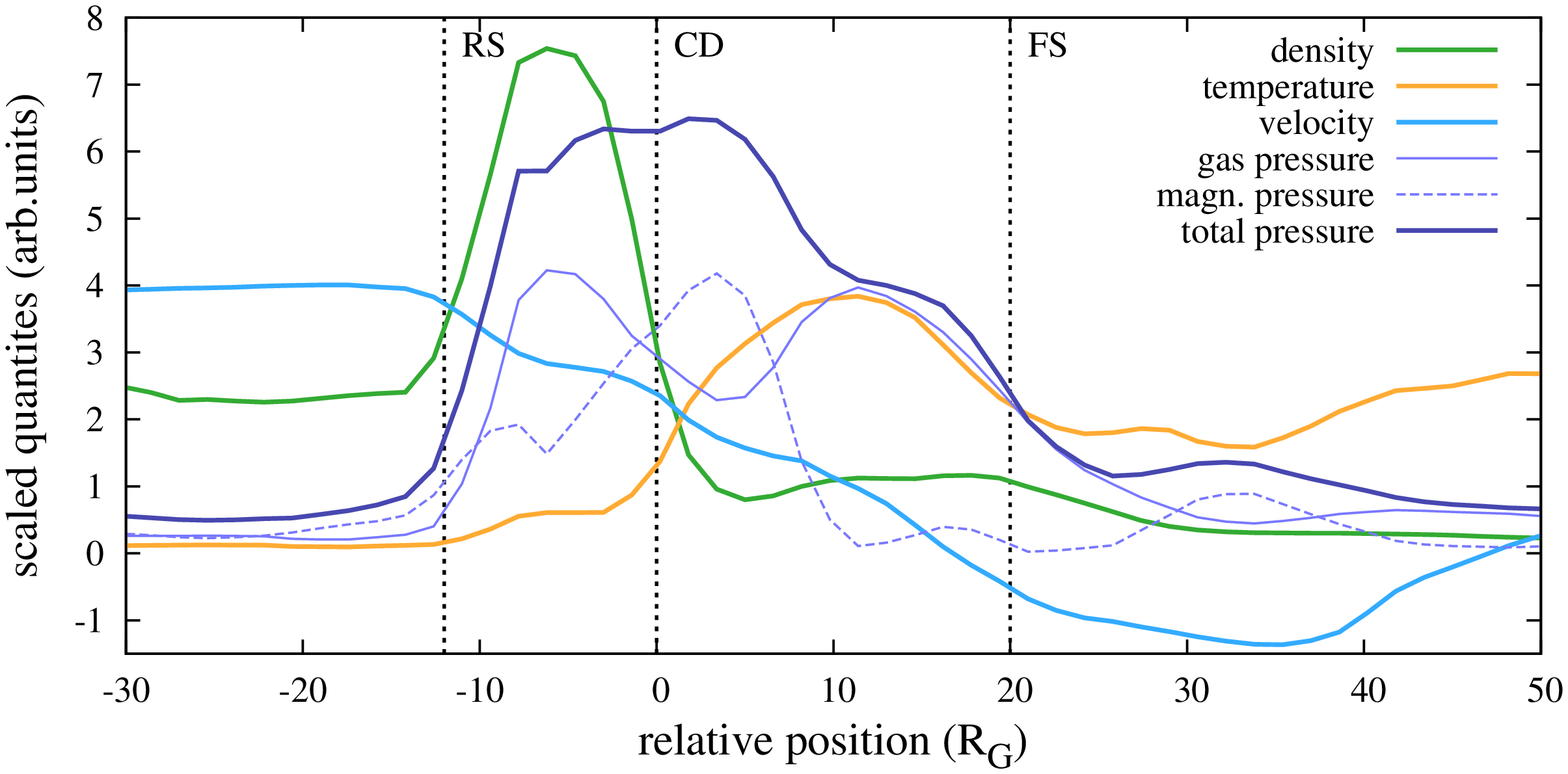}\\
\hspace{-.05cm}\includegraphics[angle=270,width=1.\columnwidth]{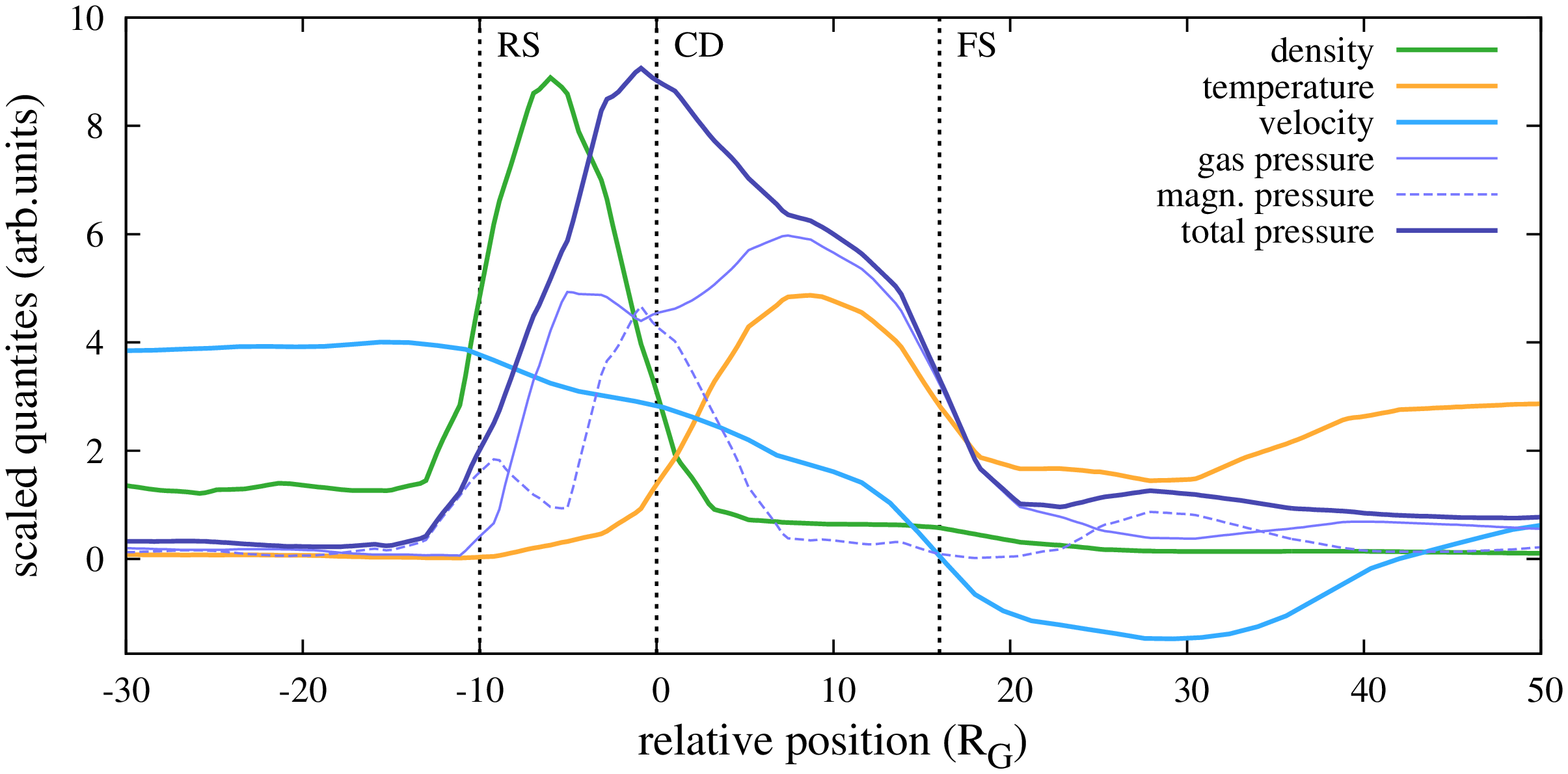}\\
\caption{Profiles of density, temperature, velocity, gas pressure, magnetic pressure, and 
total pressure across the shock for models N0 (top) and F0 (bottom) at
$t=t_0-0.63\rm yr$. The cloud front is moving to the right.  The vertical
dotted lines show approximate locations of the reverse shock (RS), contact
discontinuity (CD) and forward shock (FS).}
\label{f.lines}
\end{figure}

\begin{figure}
  \centering
\hspace{-.05cm}\includegraphics[angle=270,width=1.\columnwidth]{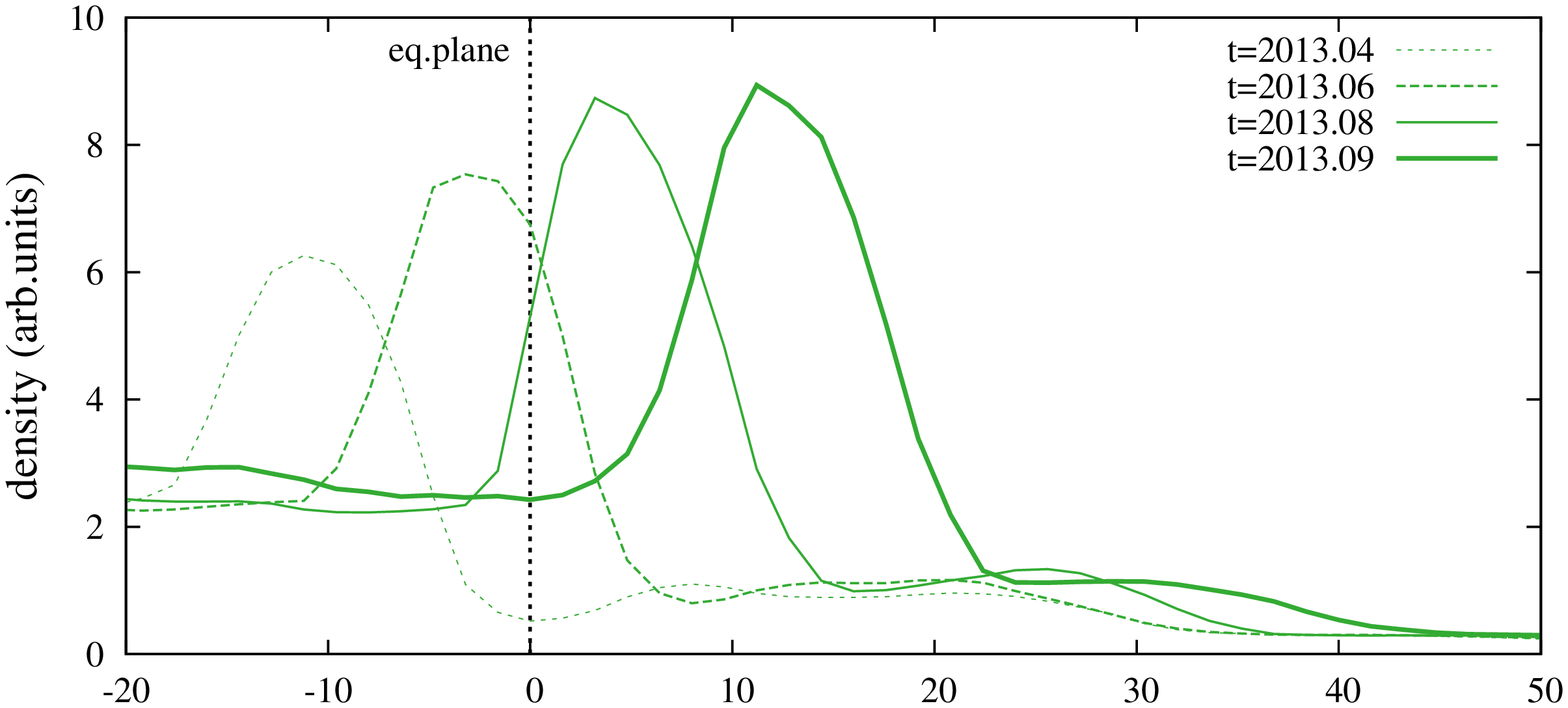}\\
\vspace{-.3cm}\hspace{-.05cm}\includegraphics[angle=270,width=1.\columnwidth]{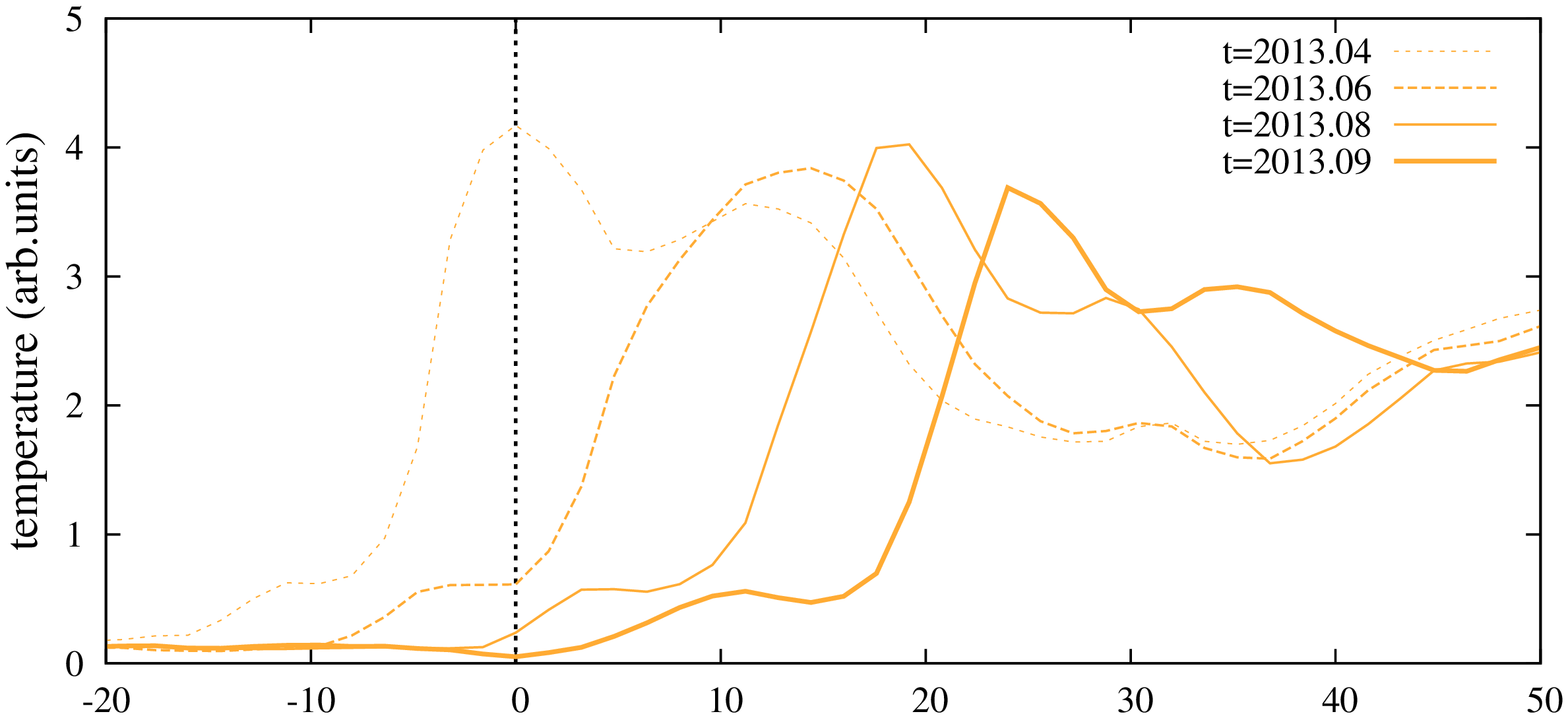}\\
\vspace{-.3cm}\hspace{-.05cm}\includegraphics[angle=270,width=1.\columnwidth]{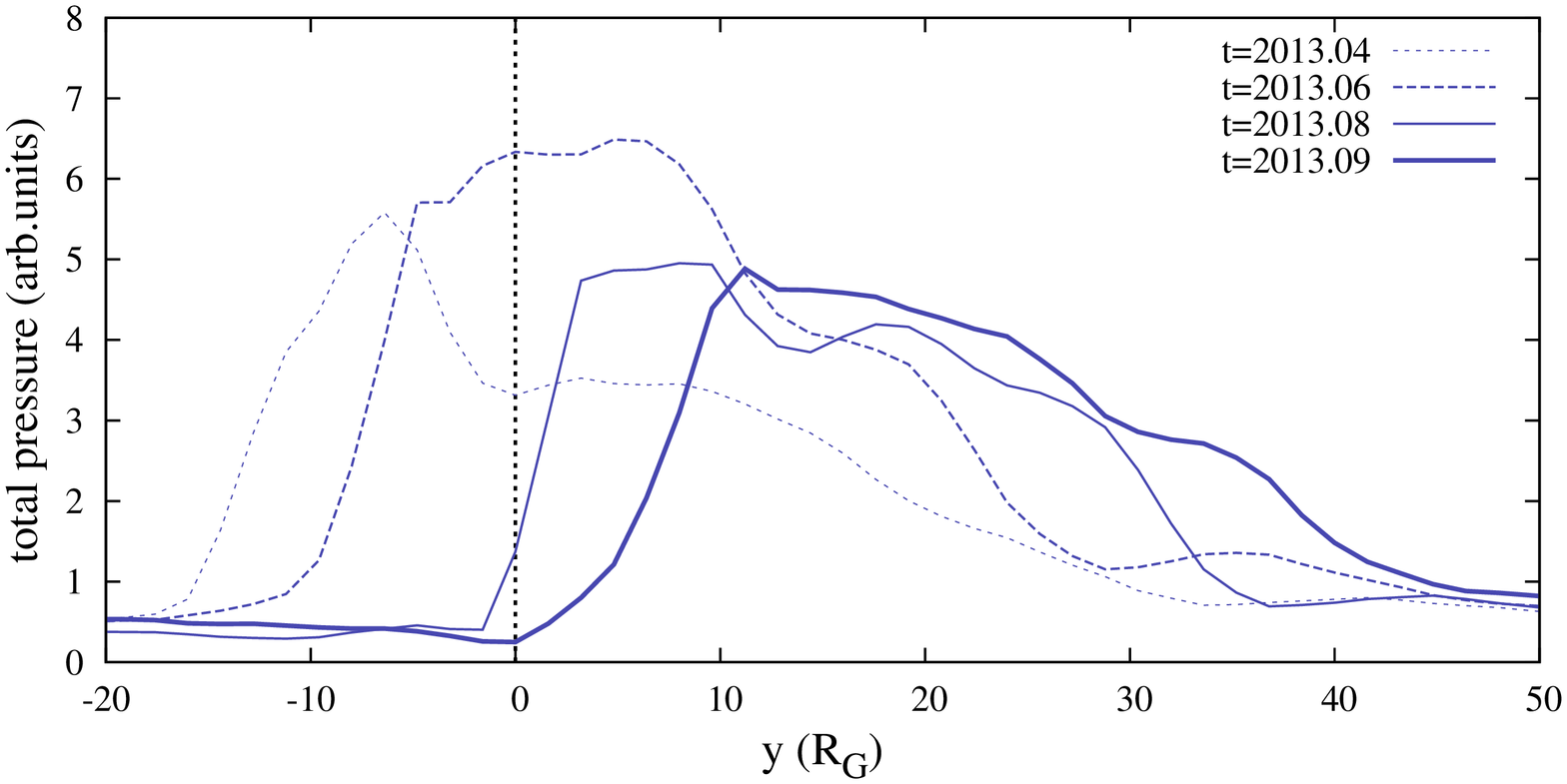}\\
\caption{Evolution of density (top), temperature (middle), and total pressure 
(bottom panel) across the shock with time for model N0. The vertical
dotted line shows the location of the disk equatorial plane.}
\label{f.cl2time}
\end{figure}

\begin{figure}
  \centering
\hspace{-.05cm}\includegraphics[angle=270,width=1.\columnwidth]{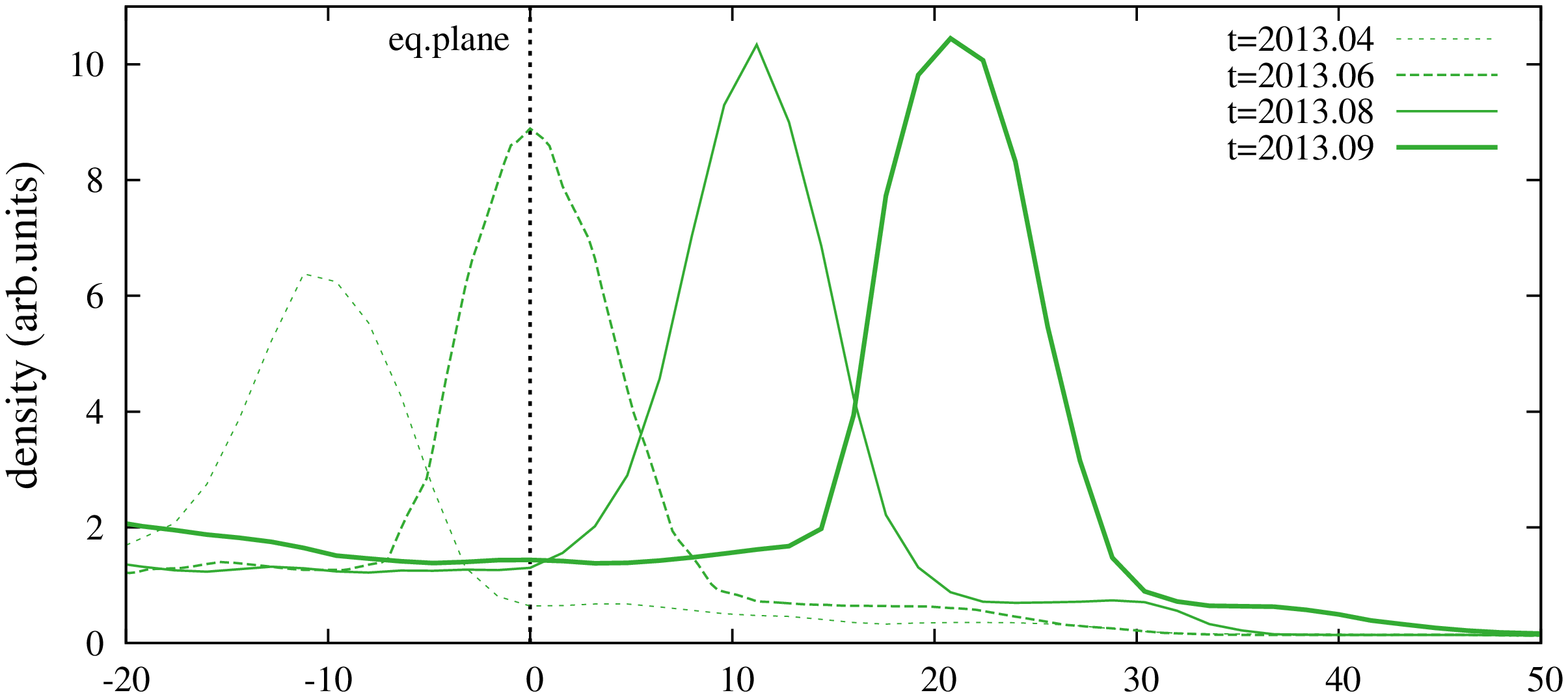}\\
\vspace{-.3cm}\hspace{-.05cm}\includegraphics[angle=270,width=1.\columnwidth]{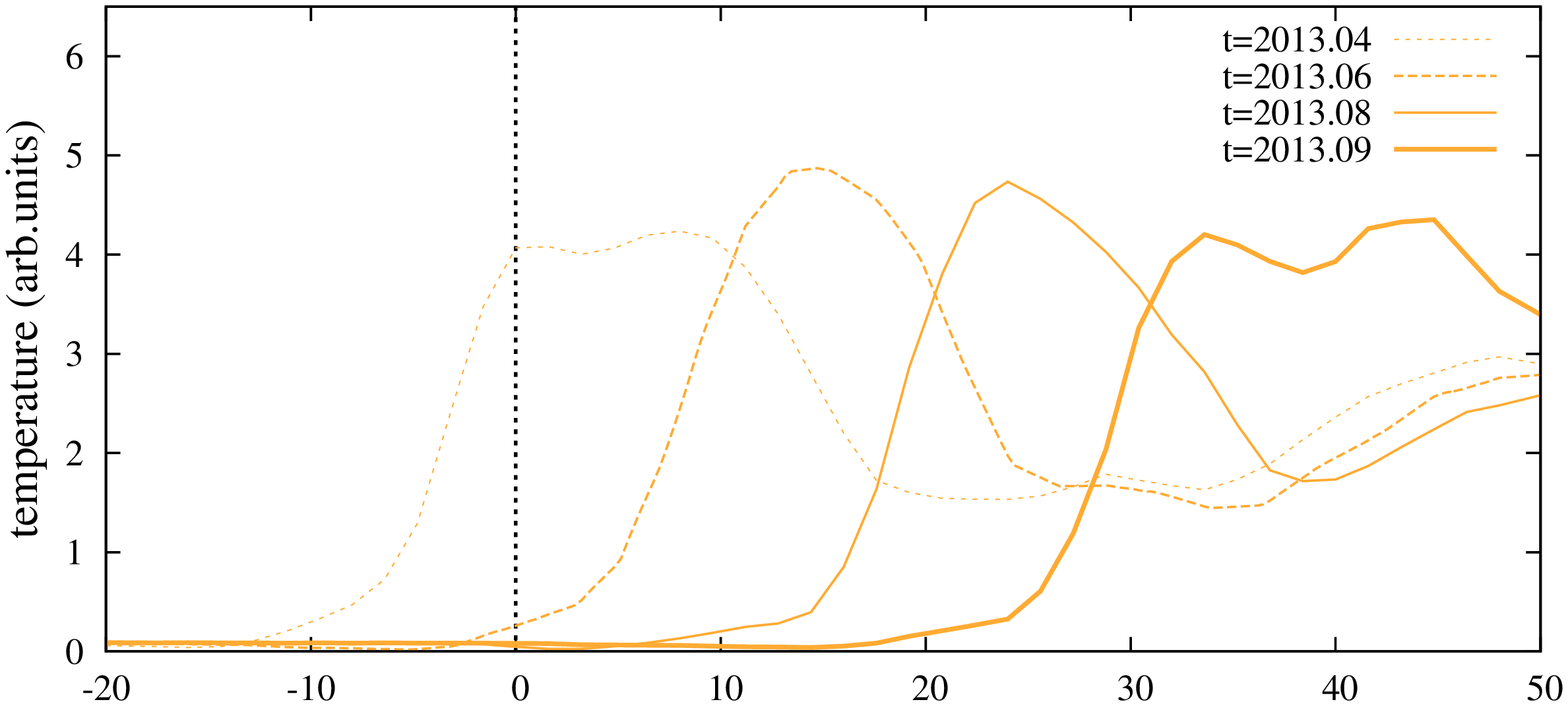}\\
\vspace{-.3cm}\hspace{-.05cm}\includegraphics[angle=270,width=1.\columnwidth]{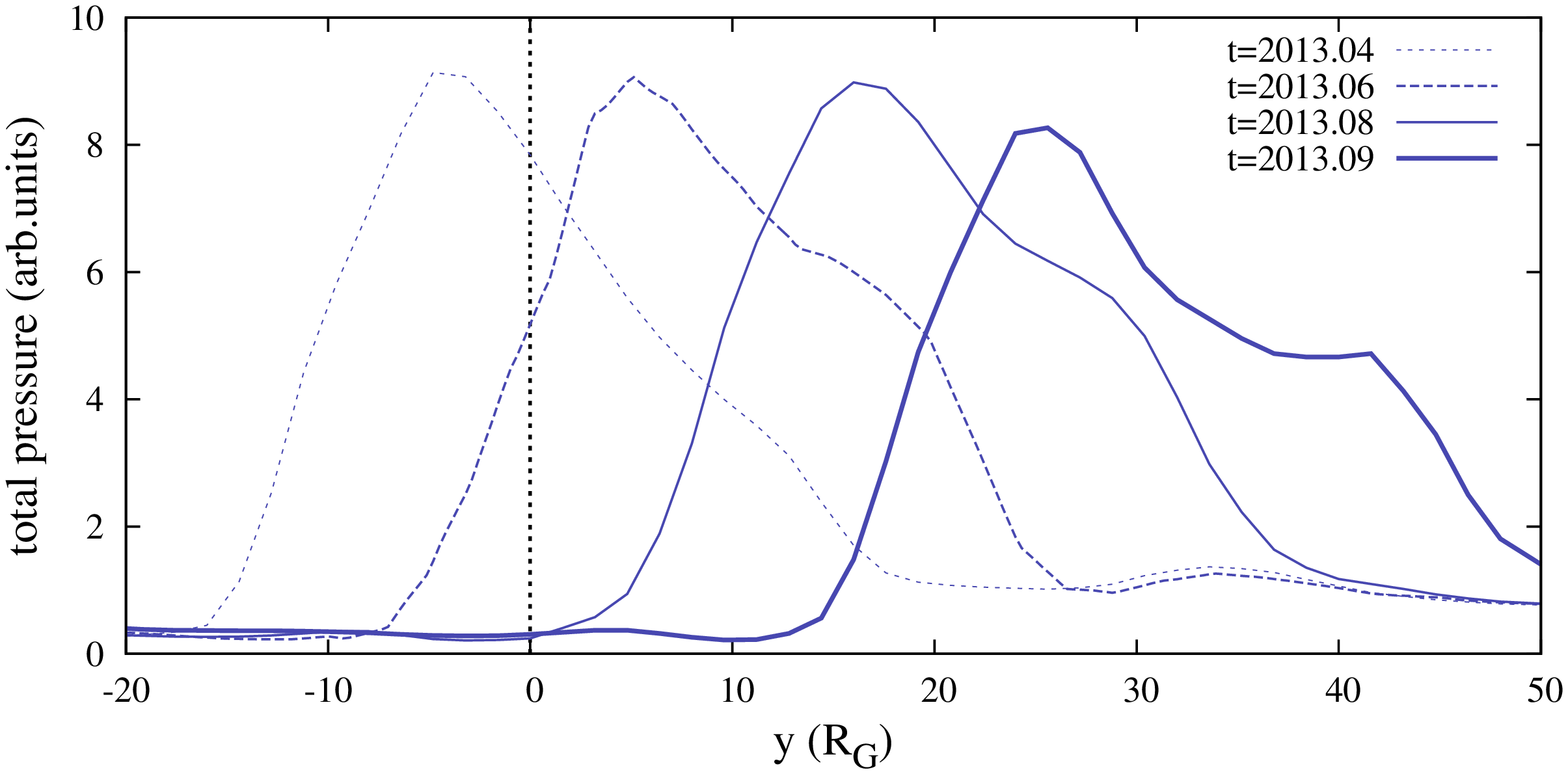}\\
\caption{Similar to Fig.~\ref{f.cl2time} but for model F0.}
\label{f.cl2flattime}
\end{figure}

We performed 12 simulations and studied in detail the
flow structure as the cloud front sweeps past pericenter.  In
Fig.~\ref{f.panels} we show orbital plane shapshots of density,
temperature, and magnetic pressure for two representative models, N0
and F0, at a time when the shock has reached the orbit pericenter. In both models the cloud orbit is
inclined at an angle $i=60\,\rm deg$ and crosses the disk equatorial
plane once ($\omega=0$).  Colors in each panel denote the magnitude of
the relevant quantity in arbitrary units, vectors show the velocity
field, yellow dashed lines show the nominal orbit of the cloud's
center-of-mass, and dashed vertical lines perpendicular to the cloud
front show the line along which profiles of various quantities are
extracted and plotted in Figs.~\ref{f.lines}, \ref{f.cl2time}, and
\ref{f.cl2flattime}.

The left panels in Fig.~\ref{f.panels} show the distribution of
density in these snapshots. The cloud has a large density and momentum
relative to the ambient gas in the accretion disk, allowing it to plow
through the accretion disk easily. However, the tidally stretched front
of the cloud is less massive in comparison and experiences
hydrodynamic and magnetic drag from the ambient gas. This interaction
sculpts the front of the cloud into an oblate shape. We can
distinguish four distinct regions in the structure, which from bottom
to top are as follows: (i) undisturbed turbulent disk gas at the
bottom (white and brown shades, compare with
Fig.~\ref{f.originalslice}), (ii) denser shocked disk gas above (light
blue) which has been processed through the forward bow-shock which
separates these two zones, (iii) high density shocked cloud gas
further up (black) which is separated from the shocked
external gas by a contact discontinuity, and (iv) unshocked cloud gas
 which lies above the reverse shock.

The front of the bow-shock is not aligned with the orbit of the cloud
center-of-mass, but lies at a substantially smaller radius,
$R=60\tilde R_{\rm G}$ instead of $R=110\tilde R_{\rm G}$. This is
because of two effects.  First, given the initial shape and velocity
distribution of the cloud, the parts of the cloud that reach
pericenter first are those that have less angular momentum and hence
pericenters closer to the black hole. Second, the interaction between
the front layers of the cloud and the disk gas causes some additional
loss of angular momentum in the cloud material
\footnote{The simulations we performed do not allow us to study
    the impact of the cloud on the BH mass accretion rate. It is
    expected that most of the additional accreted material would come
    from "filaments" stripped from the cloud surface. Yet, we likely
    underestimate this effect because of poor resolution. Furthermore, the innermost region of our simulation domain is not
    self-similar (see Section \ref{s.mhdsetup}) and therefore after
    scaling it up it will not reproduce the real Sgr A* accretion disk
    at the corresponding radii. Thus, any estimate of the cloud mass
    deposited at these radii is likely to be unreliable.}.  The effective area of the forward
bow-shock is approximately $A=\pi R_{\rm front}^2$, with $R_{\rm
  front}\approx 40\tilde R_{\rm G}= 10^{15}\rm cm$. This is consistent
with the area assumed in \cite{sadowski+G21}.

The middle panels of Fig.~\ref{f.panels} show the distribution of
temperature.  The four zones previously mentioned can also be
distinguished in the temperature maps. As expected, the
external gas is hot, the shocked external gas is even hotter, the
shocked cloud gas is warm, while the unshocked cloud gas is cold. The
shape of the bow-shock is most clearly seen in the bottom panel.

The right two panels show the distribution of magnetic pressure. As
the magnetized external gas passes through the forward shock, the
perpendicular (to the relative velocity vector) field is amplified by
compression. This field then drapes around the contact discontinuity
and contributes substantially to the total pressure in this region of
the shocked gas. This strong field likely also inhibits
Kelvin-Helmholtz instability at the contact discontinuity interface
(Markevitch \& Vikhlinin 2007).

Fig.~\ref{f.lines} shows in greater detail the one-dimensional
profiles of various quantities of interest along the vertical dashed
lines in Fig~\ref{f.panels} for models N0 (top) and F0 (bottom).  The three
dotted lines in the figure show the approximate locations of the
forward shock (FS), contact discontinuity (CD) and reverse shock (RS).
As can be seen, the shocks are not well resolved, both because the
native resolution of the simulation is somewhat low and because the
trajectory under consideration crosses the numerical grid at an angle in
the $\theta\phi$ plane.  Nevertheless, the basic features of the jumps
across FS, CD and RS are clearly present.  Across the FS, the density
(green curve) jumps from its unshocked value up by a factor of a few,
as do the temperature (orange) and total pressure (dark blue). The gas 
thermal pressure (thin blue line) dominates the pressure balance around 
the FS, whereas a region with high magnetic pressure (dashed blue line) 
is present close to the CD. The magnetic field accumulated here comes 
from the shock encounter with a region in the accretion flow where the
 magnetic pressure exceeded the gas pressure (as in the pre-shock region 
at $\sim30 \,\tilde R_{G}$ in Fig~\ref{f.lines}). The jumps across the FS are
consistent with a strong shock with Mach number of a few. More
precisely, from the density jump $\rho_d/\rho_u$ across the forward
shock, we obtain an estimate for the shock Mach number
\begin{equation}
M=\sqrt{\frac{2\, \rho_d/\rho_u}{\hat{\gamma}+1-(\hat{\gamma}-1)\, 
\rho_d/\rho_u}}\simeq 2.5-3.0
\end{equation}
where $\hat{\gamma}=5/3$ is the adiabatic index. Alternatively, from
the jump $p_d/p_u$ in the gas thermal pressure (which dominates the
pressure balance in the vicinity of the forward shock, see
Fig.~\ref{f.lines}), we derive
\begin{equation}
M=\sqrt{\frac{(\hat{\gamma}+1)\, p_d/p_u+\hat{\gamma}-1}{2 \hat{\gamma}}}\simeq2-2.5
\end{equation}
Both estimates concur to suggest a moderate value for the shock Mach
number at pericenter, $M\sim2-3$, in agreement with the assumptions of
\cite{sadowski+G21}. However, it is possible that we slightly underestimate
the shock strength because of the limited resolution and the resulting numerical diffusion.

There are somewhat larger jumps across the RS, which has a larger Mach
number since the unshocked cloud gas is very cold. In particular, the
density of the shocked gas (between the RS and the CD) is quite
large. As expected, the pressure is continuous across the CD, and
correspondingly the density and the temperature have inversely related
jumps.  

The cyan curve shows the longitudinal velocity along the cut.
The velocity goes from that of the external gas to that of the cloud
in two jumps, one each at FS and RS. In addition, between the shocks
and the CD, there is a gradient in the velocity, which we discuss
next.

While the profiles described above are qualitatively consistent with
the expectations for a planar FS--CD--RS double-shock system, the full
structure we have at hand is indeed a three-dimensional bow-shock.
This introduces some important differences, caused by the fact that,
in a bow-shock, gas can flow to the side and slide away along the
contact discontinuity \citep{landaufluid}. In order to understand the
differences specific to this geometry, we used the numerical code
\texttt{KORAL} \citep{sadowski+koral} to simulate a toy hydrodynamical
problem in which a solid spherical object plows through a homogeneous
hot fluid with a relative Mach number ${\cal M}=2.5$.

\begin{figure}
  \centering
\hspace{0cm}\includegraphics[width=1.\columnwidth]{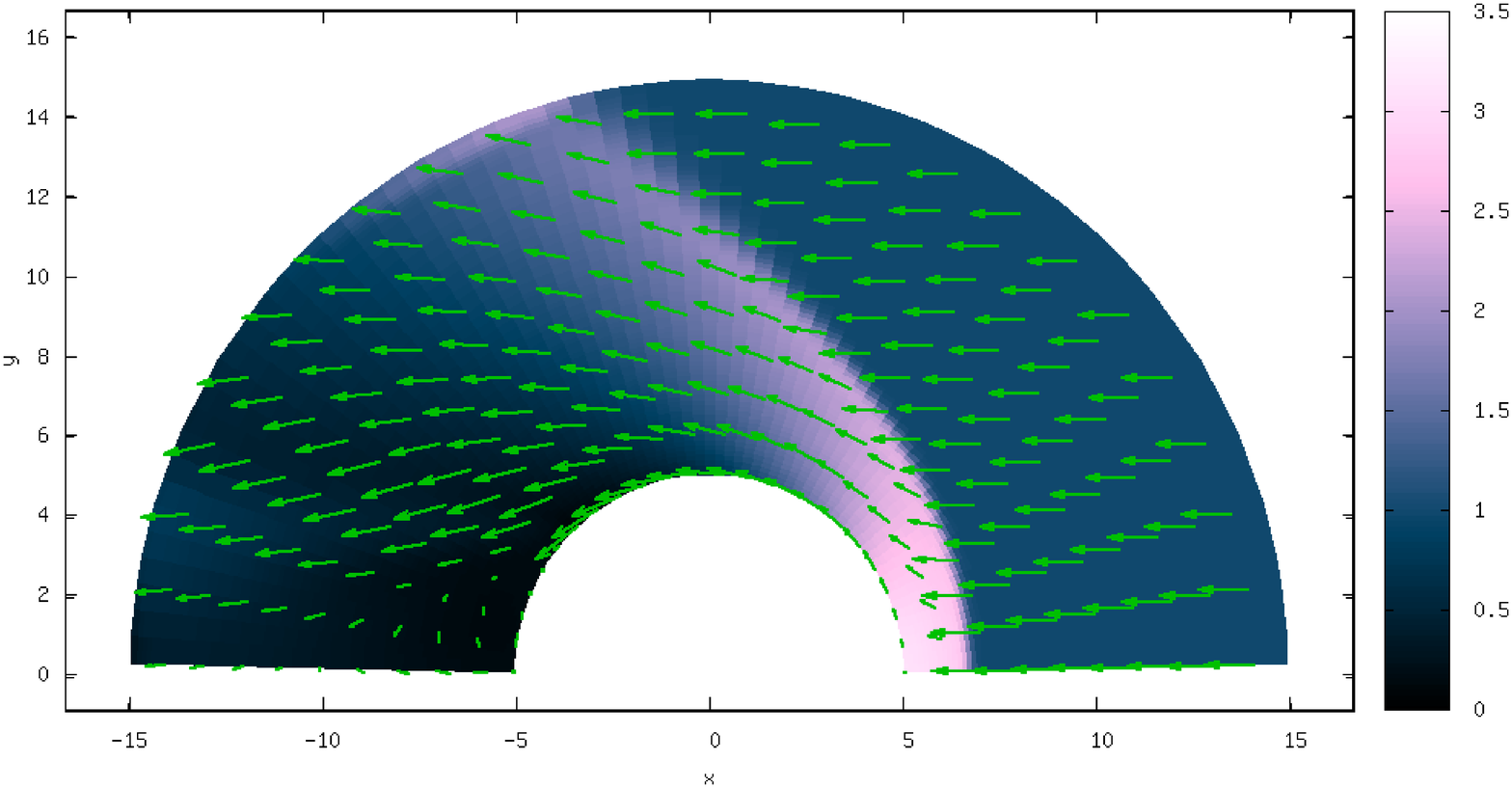}\\
\hspace{-.05cm}\includegraphics[angle=270,width=1.0\columnwidth]{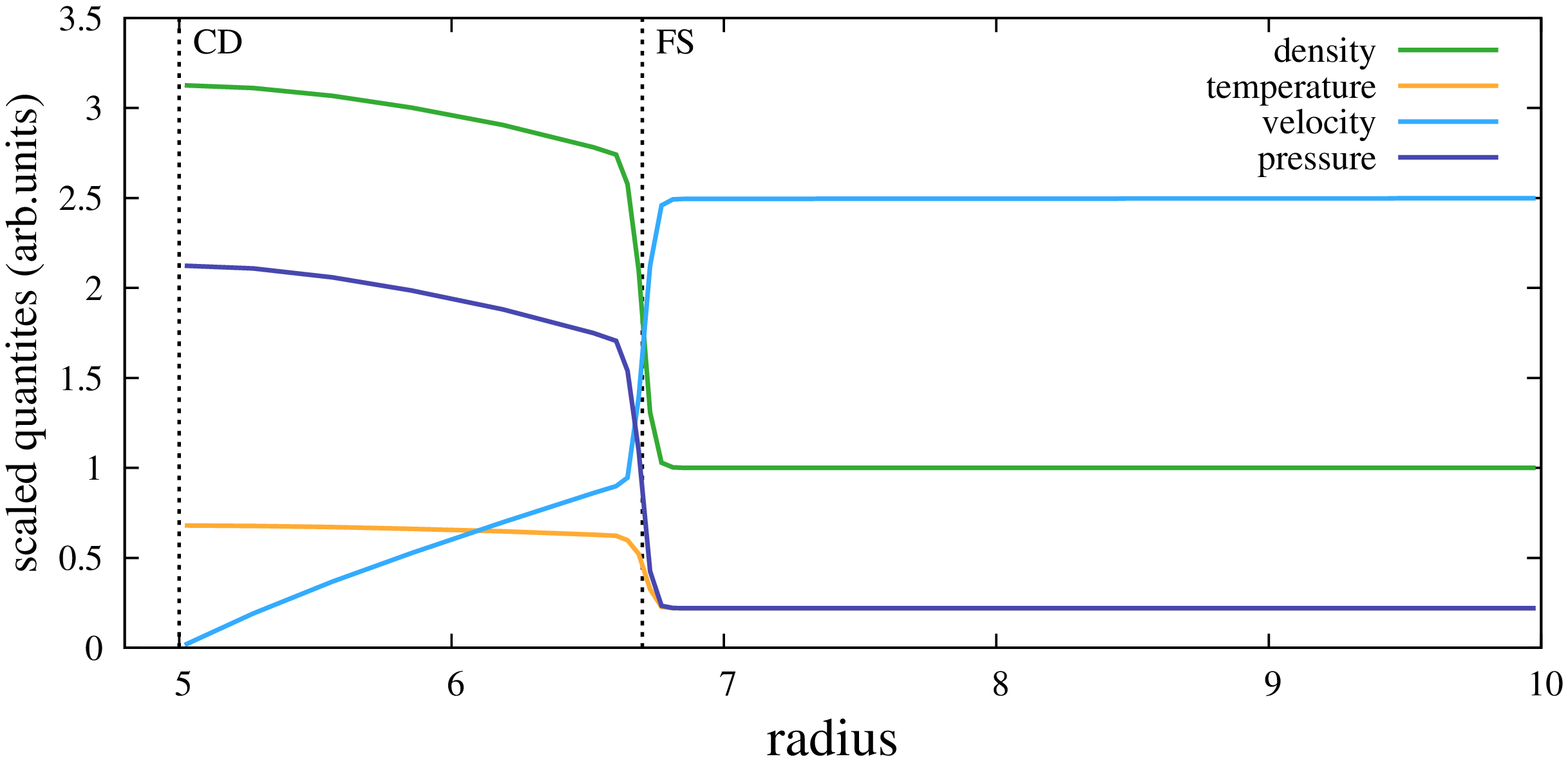}\\
\caption{Top: Density profile and the velocity field in a toy three-dimensional 
bow shock simulation with ${\cal M}=2.5$ in spherical coordinates. Bottom:
Profiles of density, pressure, temperature and velocity along the
impact axis.  }
\label{f.bow3d}
\end{figure}

Fig.~\ref{f.bow3d} shows the results from this hydrodynamic
simulation after the system has reached steady state. Because we assume a
rigid sphere, there is no reverse shock, and the surface of the sphere
plays the role of the contact discontinuity. The forward bow-shock,
however, is very clearly delineated in the top panel. The bottom panel
shows profiles of density, pressure, temperature and velocity along
the symmetry axis. The jumps in these quantities at the FS are as
expected from standard theory of adiabatic shocks. The main
differences from the planar shock problem appear in the post-shock
gas. Whereas for the planar case we expect all gas properties to be
independent of position after the shock, in the bow-shock case we see
that the density, temperature and pressure all increase between the FS
and the CD, whereas the velocity drops down to zero at the CD.  These
differences are caused by gas being diverted around the CD, as seen from
the velocity vectors in the top panel.

A comparison of the toy model results in Fig.~\ref{f.bow3d} with the
G2 simulation results in Fig.~\ref{f.lines} indicates that similar
variations are present in the latter.  The variations in the pressure
and velocity, in particular, are quite similar to those seen in the 3D
hydrodynamical problem, suggesting that the shock geometry is
responsible for the behavior seen in Fig.~\ref{f.bow3d}.

To investigate the effects of the magnetic structure of the accretion flow on the cloud-disk encounter, we have run the simulation N0B, without magnetic fields. As compared to our fiducial model N0, we find that the run N0B yields a similar structure for the FS, which is consistent with the fact that pressure balance at the FS is dominated by the gas thermal component (see Fig~\ref{f.lines}). However, the RS in N0B is weaker than in N0, lacking the magnetic support present for N0 in the vicinity of the CD.

We also study the time evolution of the shock by looking at the
shock structure at different snapshots in the simulation.
Figs.~\ref{f.cl2time} and \ref{f.cl2flattime} show cuts across the
FS--CD--RS structure for models N0 and F0, respectively, at four different times. The cloud
moves from left to right and the vertical dotted line indicates the
location of the disk mid-plane. The various features discussed earlier
are clearly seen in all the snapshots and the whole pattern moves
steadily to the right. In addition, there is evidence for some growth 
in the amount of shocked gas with increasing time.

\subsection{Pericenter crossing times }

\cite{sadowski+G21} calculated synchrotron emission from relativistic 
electrons accelerated in the bow shock of G2 and showed that the
maximum radio synchrotron emission is expected roughly a month after
the bow shock crosses pericenter. It is evident from the cloud-disk
interaction models we have presented here that the bow shock forms ahead of
the front of the cloud, and reaches pericenter well ahead of the
pericentric passage of the cloud's center-of-mass.  We list in
the last column of Table~\ref{t.models} the times when the cloud front reaches the location
corresponding to true anomaly $\theta=0$ of the center-of-mass orbit
in each of our 12 simulations. In
most models the crossing occurs 7 to 9 months before the cloud
  center-of-mass'
  epoch of pericenter ($t_0$).  A
larger initial cloud size (models N2 and F2) causes the crossing to
occur two or three months earlier. On the other hand, clouds on orbits
with low inclinations (N0LN and N0LC) or with double crossing (N0D)
are slowed down by interaction with the disk gas compared to the
fiducial model N0. In these models, the cloud front reaches pericenter
roughly a month later. Model N0B with no magnetic fields shows an
early crossing time by about a month, indicating that drag due to the
magnetic field draped around the CD plays a dynamically significant
role in all the other models.

\section{Discussion}
\label{s.discussion}

In this paper, we performed detailed magnetohydrodynamic simulations
of the interaction between the gas cloud G2 and the accretion flow
around the Galactic Center black hole Sgr~A$^*$. We scaled down the cloud
to fit it within the converged region of previously simulated accretion
flow model solved on a fixed numerical grid. Despite the limited numerical resolution
we were able to simulate the formation of the bow-shock region and qualitatively
study its properties.
We showed that for any
reasonable cloud structure and orbit orientation that fit the current
data, the shock pericenter crossing takes place approximately 7 to
  9 months
 earlier than the epoch when the cloud center-of-mass reaches
pericenter. Depending on the orbital model, we predict the peak
  radio emission in Spring 2013 for \cite{gillessen+12b} or late Summer 2013
for \cite{keckg2} orbit solutions, respectively.

The Very Large Array (VLA) carried out observations
of the Galactic center during
October 2012, December 2012, February 2013, and March 2013 as part of a monitoring
campaign.\footnote{Data are publicly available at\newline
\texttt{https://science.nrao.edu/science/service-observing}.}  The
data taken in February and March provide only upper limits to the flux from
\sgra\ at low frequencies because the VLA was in the low-resolution
D-configuration during this observation and could not fully disentangle
the fluxes of Sgr A* from that of other neraby sources. Intriguingly, the flux at 14 GHz,
which suffers less from source confusion, shows a modest flux
increase, which could be due to the extra radio emission from the bow
shock or due to intrinsic variability of the source.Another
observation was made by \cite{kassim+atel} using
the Giant Metrewave Radio Telescope (GMRT) in late January and early
February 2013 and resulted in non-detection of sub-GHz emission.


February and March monitoring observations did not show a brightening
but there are recent reports of brightening at 22GHz and 32 GHz
\citep{brunthaler+atel, tsuboi+atel}. It remains to be seen whether
this flux enhancement is due to the bow shock.

We encourage continued monitoring of the
Galactic Center in the very near future. Higher resolution
observations can not only verify a flux enhancement but may also be
able to detect the displacement of the known radio source coincident
with the black hole from the bow shock emission $\sim 4400 R_{\rm G}$ away.
Finally, observations at higher frequencies, even with the more
compact array configurations, could differentiate an intrinsic
brightening. Even though we predict the largest flux increase around a
GHz, significant enhancement can be observed up to $\sim 14$~GHz.

\section{Acknowledgements}

A.S. and R.N. were supported in part by NASA grant NNX11AE16G. L.S.
is supported by NASA through Einstein Postdoctoral Fellowship grant
number PF1-120090 awarded by the Chandra X-ray Center, which is
operated by the Smithsonian Astrophysical Observatory for NASA under
contract NAS8-03060. F.\"O. acknowledges support from NSF grant
AST-1108753 and from the Radcliffe Institute for Advanced Study at
Harvard University.  The simulations were performed on XSEDE resources
under contract No. TG-AST120010, and on NASA High-End Computing (HEC)
resources through the NASA Advanced Supercomputing (NAS) Division at
Ames Research Center.
 
\bibliographystyle{mn2e}

{\small
\begin{thebibliography}{}

\bibitem[Anninos et al.(2012)]{anninos+12} Anninos, P., Fragile, 
P.~C., Wilson, J., \& Murray, S.~D.\ 2012, \apj, 759, 132 


\bibitem[Baganoff et al.(2001)]{2001Natur.413...45B} Baganoff, F.~K., 
Bautz, M.~W., Brandt, W.~N., et al.\ 2001, \nat, 413, 45 

\bibitem[Broderick 
\& Loeb(2005)]{2005MNRAS.363..353B} Broderick, A.~E., \& Loeb, A.\ 2005, \mnras, 363, 353 

\bibitem[Broderick et al.(2011)]{broderick+11} Broderick, A.~E., 
Fish, V.~L., Doeleman, S.~S., \& Loeb, A.\ 2011, \apj, 735, 110 

\bibitem[Brunthaler et al.(2013)]{brunthaler+atel} Brunthaler A., Falcke
  F., Bower G. C. , Ott J., Reid M. J.\ 2013, The
Astronomer's Telegram, 5014, 1

\bibitem[Burkert et al.(2012)]{burkert+12} Burkert, A., 
Schartmann, M., Alig, C., Gillessen, S., Genzel, R., Fritz, T.~K., 
\& Eisenhauer, F.\ 2012,  \apj, 750, 58 

\bibitem[Dexter et al.(2010)]{dexter+10} Dexter, J., Agol, E., 
Fragile, P.~C., \& McKinney, J.~C.\ 2010, \apj, 717, 1092 

\bibitem[Dibi et al.(2012)]{dibi+12} Dibi, S., Drappeau, S., 
Fragile, P.~C., Markoff, S., \& Dexter, J.\ 2012, \mnras, 426, 1928 

\bibitem[Gammie et al.(2003)]{gammieetal03} Gammie, C.~F., McKinney, 
J.~C., \& T{\'o}th, G.\ 2003, \apj, 589, 444 


\bibitem[Gillessen et al.(2012)]{gillessen+12a} Gillessen, S., 
Genzel, R., Fritz, T.~K., et al.\ 2012a, \nat, 481, 51 

\bibitem[Gillessen et al.(2013)]{gillessen+12b} Gillessen, S., 
Genzel, R., Fritz, T.~K., et al.\ 2013, \apj, 763, 78 

\bibitem[Kassim et al.(2013)]{kassim+atel} Kassim, N.~E., Hyman,
  S.~D., Intema, H., Clarke, T.~E., \& Subrahmanyan, R. 2013, The
Astronomer's Telegram, 4922, 1

\bibitem[Landau 
\& Lifshitz(1959)]{landaufluid} Landau, L.~D., \& Lifshitz, E.~M.\ 1959, Course of theoretical physics, Oxford: Pergamon Press, 1959,  

\bibitem[Markevitch 
\& Vikhlinin(2007)]{2007PhR...443....1M} Markevitch, M., \& Vikhlinin, A.\ 2007, Physics Reports, 443, 1 




\bibitem[Marrone et al.(2007)]{marrone07} Marrone, D.~P., Moran, 
J.~M., Zhao, J.-H., \& Rao, R.\ 2007, \apjl, 654, L57 


\bibitem[Mo{\'s}cibrodzka et al.(2009)]{moscibrodzka+09} 
Mo{\'s}cibrodzka, M., Gammie, C.~F., Dolence, J.~C., Shiokawa, H., 
\& Leung, P.~K.\ 2009, \apj, 706, 497 


\bibitem[Narayan et al.(2003)]{2003PASJ...55L..69N} Narayan, R., 
Igumenshchev, I.~V., \& Abramowicz, M.~A.\ 2003, \pasj, 55, L69 


\bibitem[Narayan et al.(2012a)]{narayan+12a} Narayan, R., {\"O}zel, 
F., \& Sironi, L.\ 2012a, \apjl, 757, L20 

\bibitem[Narayan et al.(2012b)]{narayan+12b} Narayan, R., Sadowski, 
A., Penna, R.~F., \& Kulkarni, A.~K.\ 2012b, MNRAS, accepted

\bibitem[{Penna} et~al.(2012){Penna}, {Kulkarni} \& {Narayan}]{penna+torus}
{Penna} R., {Kulkarni} A., {Narayan} R., 2012, Submitted to \aap

\bibitem[Phifer et al.(2013)]{keckg2} Phifer, K., Do, T., 
Meyer, L., Ghez, A.~M., Witzel, G., Yelda, S., Boehle, A., Lu, J.~R., 
Morris, M.~R., Becklin, E.~E., \& Matthews, K.\ 2013,  arXiv:1304.5280 


\bibitem[Psaltis(2012)]{2012ApJ...759..130P} Psaltis, D.\ 2012, \apj, 759, 
130 

\bibitem[Psaltis et al.(2013)]{ps13} Psaltis, D., Narayan, R., \& Broderick A.\ 2013, in preparation

\bibitem[Saitoh et al.(2012)]{saitoh+12} Saitoh, T.~R., Makino, 
J., Asaki, Y., et al.\ 2012, arXiv:1212.0349 

\bibitem[S{\k a}dowski et al.(2013a)]{sadowski+koral} S{\k a}dowski, 
A., Narayan, R., Tchekhovskoy, A., \& Zhu, Y.\ 2013a, \mnras, 429, 3533 

\bibitem[S{\k a}dowski et al.(2013b)]{sadowski+G21} S{\k a}dowski, A., Sironi, 
L., Abarca, D., et al.\ 2013b, arXiv:1301.3906 

\bibitem[Schartmann et al.(2012)]{shartmann+12} Schartmann, M., 
Burkert, A., Alig, C., et al.\ 2012, \apj, 755, 155 

\bibitem[Shcherbakov et al.(2012)]{roman+12} Shcherbakov, R.~V., 
Penna, R.~F., \& McKinney, J.~C.\ 2012, \apj, 755, 133 

\bibitem[Tsuboi et al.(2013)]{tsuboi+atel} Tsuboi, M., Asaki, Y.,
  Yonekura Y., Kaneko H., Miyamoto, Y., Seta M., Nakai N., Kameya,
  O. et al.\ 2013, The
Astronomer's Telegram, 5013, 1



\bibitem[Yuan et al.(2003)]{yuan+03} Yuan, F., Quataert, E., 
\& Narayan, R.\ 2003, \apj, 598, 301 



\end{thebibliography}
}

\end{document}